\documentclass[12pt]{article}
\makeatletter

\makeatletter
\def\@cite#1{$^{\mbox{\scriptsize{#1}}}$}
\def\@biblabel#1{(#1)}
\makeatother
 
\usepackage{graphicx}
\usepackage{float}
\restylefloat{figure}

\title{\LARGE Epitaxial Transition from Gyroid to Cylinder in a Diblock Copolymer Melt}

\author{
Takashi Honda
\thanks{Japan Chemical Innovation Institute, 
and Division of Polymer, Faculty of Engineering, 
Tokyo Institute of Technology,
Ookayama, Meguro-ku,
Tokyo 152-8552, Japan}
\and
Toshihiro Kawakatsu
\thanks{Department of Physics, Tohoku University,
Aoba, Aramaki, Aoba-ku, Sendai 980-8578, Japan}
}

\date{September 23, 2005}

\begin{document} 
\maketitle





{\Large\bf Abstract} \\

An epitaxial transition from a bicontinious double gyroid to a hexagonally packed cylinder structure induced by an external flow is simulated using real-space dynamical self-consistent field technique.  In order to simulate the structural change correctly, we introduce a system size optimization technique by which emergence of artificial intermediate structures are suppressed.  When a shear flow in [111] direction of the gyroid unit cell is imposed, a nucleation and growth of the cylinder domains is observed.  We confirm that the generated cylindrical domains grow epitaxially to the original gyroid domains as gyroid $d_{\{220\}}$ $\rightarrow$ cylinder $d_{\{10\}}$.
In a steady state under the shear flow, the gyroid shows different reconnection processes depending on the direction of the velocity gradient of the shear flow.  A kinetic pathway previously predicted using the self-consistent field theory where three fold junctions transform into five fold junctions as an intermediate state is not observed. 

\section{Introduction}

In the past, the self-organized microdomain structures of diblock copolymers have been the target of extensive studies  
\cite{Matsen Bates,Hamley book,Bates Fredrickson,Fredrickson}.
Especially, the order-order transitions (OOTs) between the microdomain structures are one of the central issues of the current experimental and theoretical studies. 
Among various microdomain structures, the bicontinuous double gyroid (G) structure has attracted a great interest because of its complex structure (space group $Ia\bar{3}d$)
\cite{Hajduk}.  
Although this G phase exists only in a narrow region of the phase diagram of a diblock copolymer, i.e. the region between the lamellar (L) phase and the hexagonally packed cylinder (C) phase
\cite{Matsen Bates 96},
its complex domains are expected to have a wide applicability to various techniques, for example, microporous systems, nano-reactors, and so on
\cite{Hashimoto,Zhao,Chan}.

Experimentally, the OOT from the G phase to the C phase is believed to be an epitaxial transition where the created cylindrical domains are commensurate to the original gyroid domains. However, the microscopic detailed process of this epitaxial transition has not been understood yet.

Epitaxial relationships between the G and C structures were observed in several experiments.
Upon a temperature change, Ran\c{c}on and Charvolin have observed that the \{10\} plane of the C is in commensurate to the \{211\} plane of the G domains in a surfactant system
\cite{Y. Rancon J. Charvolin}.
In the present paper, we will use simple notation as C \{10\} $\rightarrow$ G \{211\} for such an epitaxial relationships between planes.

An external shear flow also accelerates the OOTs. 
Under a shear flow and a temperature change, 
Schulz et al. have found different epitaxial relationships C \{10\} $\rightarrow$ G \{220\} and C \{11\} $\rightarrow$ G \{211\} in a block copolymer mixture by small-angle neutron scattering (SANS) experiments 
\cite{M. F. Schulz F. S. Bates K. Almdal K. Mortensen}. 
Under similar experimental conditions using small-angle X-ray scattering (SAXS) diffraction techniques, F\"orster et al. have observed the epitaxial relationship C \{10\} $\rightarrow$ G \{211\} similar to the Ran\c{c}on and Charvolin's observation  
\cite{Forster}.
The same epitaxial relationship was also observed by Vigild et al. in a block copolymer system using SANS
\cite{M. E. Vigild}.
A cyclic transition C $\rightarrow$ G $\rightarrow$ C in a block copolymer solution has been studied by Wang and Lodge, who supported the epitaxial relationship G \{211\} $\rightarrow$ C \{10\}
\cite{C. -Y. Wang T. P. Lodge}.

From the point of view of kinetics,
a long-lived coexistence between the C phase and the G phase has been found in the C $\rightarrow$ G of a block copolymer system under a shear flow and a temperature change
\cite{Floudas Ulrich Wiesner Chu}.
Furthermore, a grain boundary between the C phase and the G phase has been observed by a polarized optical microscopy in a quenched polymer solution
\cite{T. Q. Chastek T. P. Lodge}.
These observations suggest an existence of a stable boundary between the C phase and the G phase.

On the theoretical side, mean field theories have been used to investigate microdomain structures of diblock copolymers.
Using the self-consistent field (SCF) technique, Helfand and Wasserman have evaluated the free energy and predicted the equilibrium domain sizes of the classical phases in the strong segregation regime such as the body centered cubic crystal of spherical domains (BCC), the C phase, and the lamellar (L) phase
\cite{Helfand Wasserman 4,Helfand Wasserman 5, Helfand Wasserman 6}.
On the other hand, the phase diagram of diblock copolymer in the weak segregation regime was predicted by Leibler using the random phase approximation (RPA)
\cite{Leibler}.
Leibler's phase diagram is composed of classical phases and the disordered (D) phase depending on the values of the block ratio and the $\chi N$, i.e. the product of the Flory-Huggins interaction parameter $\chi$ and the total degree of polymerization of diblock copolymer $N$.
The entire phase diagram including both the weak segregation regime and the strong segregation regime has been constructed by Matsen and Shick using the SCF technique in the reciprocal lattice space.  Besides the classical phases, they predicted the complex G phase in the weak and intermediate segregation regime
\cite{Matsen Bates,Matsen Schick}.
This theoretical phase diagram was confirmed experimentally
\cite{Khandpur et al.}.

Despite the success of the mean field theories on the equilibrium phase behavior, the investigation on the dynamic properties has not been fully developed yet.
There have been a few trials on the dynamics of OOTs and order-disorder transitions (ODTs) of the microdomain structures of diblock copolymers using the mean field approximation.
A time dependent Ginzburg-Landau (TDGL) model was used to investigate the instability in the OOTs and ODTs such as OOTs L $\rightarrow$ C and L $\rightarrow$ S, and C $\rightarrow$ S.
\cite{Qi Wang 1,Qi Wang 2,Qi Wang 3}.
In these studies, the authors retained the most unstable modes in the Fourier amplitudes of the density fluctuations emerging in the vicinity of the critical point.
A TDGL model described in terms of Fourier modes with two sets of wave vectors with different magnitudes has been used to study the transitions D $\rightarrow$ G, G $\rightarrow$ C, and so on
\cite{Nonomura Ohta 1,Nonomura Ohta 2,Nonomura Ohta 3}.
Although the TDGL theory is efficient in investigating large-scale systems, it is in principle applicable only to the weak segregation regime.

On the other hand, the SCF theory can be used to study the phase transitions in weak, intermediate and strong segregation regimes.  The quantitative accuracy of the SCF theory is another advantage compared to the TDGL theory.  This is because the SCF theory takes the conformational entropy of the polymer chains into account precisely
\cite{Helfand Wasserman 4,Hong,Fleer,kawakatsu book}.
Using the SCF theory, Laradji et al. have investigated the epitaxial transitions such as L $\leftrightarrow$ C, C $\leftrightarrow$ S, and G $\rightarrow$ C taking the anisotropic fluctuations into account
\cite{Laradji,MatsenComment}.
Matsen has also studied the transitions C $\leftrightarrow$ S and C $\leftrightarrow$ G using the SCF theory and has proposed a nucleation and growth model of the epitaxial transitions
\cite{Matsen Cylinder,Matsen Gyroid}.

All of these theoretical studies mentioned above rely on the reciprocal space representations 
\cite{Nonomura Ohta 3,Laradji,Matsen Gyroid}
and most of experimental studies
\cite{Y. Rancon J. Charvolin,Forster,M. E. Vigild}
have supported the existence of the epitaxial OOT G \{211\} $\leftrightarrow$ C \{10\}
except for Schulz et al., who have supported the epitaxial OOT G \{220\} $\leftrightarrow$ C \{10\}.
Experimentally, the epitaxial OOT G \{211\} $\leftrightarrow$ C \{10\} and G \{220\} $\leftrightarrow$ C \{10\} are recognized as the same epitaxial relationships
\cite{M. E. Vigild}
because the diffraction peaks from both G \{211\} and G \{220\} match well with the diffraction peaks from the C \{10\}.
In this argument, however, the kinetic pathway in real space was not considered.
In Figure 1, we show a projection of the G structure onto the [111] direction in the real space.
The epitaxial relations G \{211\} $\leftrightarrow$ C \{10\} and G \{220\} $\leftrightarrow$ C \{10\} are shown in Figure 1(a) and 1(b), respectively, where the spacing of the G planes and the epitaxial cylindrical domains are shown.
As the directions of these two planes G $\{112\}$ and G $\{220\}$ are perpendicular with each other and the spacings between adjacent planes are also different, the two growth mechanisms shown in Figures 1(a) and (b) should be regarded as different ones.

Furthermore, the OOTs of block copolymer melts are first order phase transitions and the nucleation and growth process of domains is expected.  
Since these process is spatially inhomogeneous, such a transition is not compatible with the treatment in the reciprocal lattice space (Fourier space), where spatially periodic lattice structures are assumed.
Therefore, dynamical simulations in real space, such as the dynamical SCF simulation, is necessary to correctly investigate this transition
\cite{Fraaije,Zvelindovsky,Hasegawa Doi,Hamley latest}.

In the present paper, we study the epitaxial OOT G $\rightarrow$ C using the dynamical SCF theory under shear flows.  In order to treat the first order transition, we introduce a system size optimization (SSO) method, in which the side lengths of the simulation box are automatically adjusted so that the size and the shape of the simulation box are fit for the lattice spacing and the lattice axes of the ordered structures.
Recently, Barrat et al. proposed a similar technique to study equilibrium domain morphology of block copolymer systems
\cite{Barrat}.
In the course of the transition, we observe a complex transient state that is composed of cylindrical domains parallel to the G \{220\} plane. We also confirm that our SSO method can reproduce spatially inhomogeneous nucleation and growth processes. Actually we observe the coexistence between the G phase and the C phase, which is consistent with the experimentally observed first order transition behavior
\cite{T. Q. Chastek T. P. Lodge}. 
Furthermore we found that the G structure shows different deformation behaviours depending on the direction of the velocity gradient of the shear flow.

We cannot confirm the scenario of the transition proposed by Matsen
\cite{Matsen Gyroid},
where the three fold junctions transform into five fold junctions.

Finally, we clarify the kinetic pathway from the G phase to the C phase under a shear flow. To the best of our knowledge, this kinetic pathway in the real space has not been reported in the literature.


\section {Theory} 

\subsection{Dynamical self-consistent field theory}

Here, we briefly summarize the SCF theory for an A-B diblock copolymer
\cite{Fredrickson,Fleer,Fraaije,Morita Kawakatsu}.
Let us consider a melt of A-B diblock copolymer.  
Due to the screening effect in the melts, we can assume Gaussian statistics for the chain conformation.
Within this Gaussian statistics, the $K$-type ($K=$A or B) segment is characterized by the effective bond length $b_K$, and the $K$-type block is characterized by the degree of polymerization $N_K$.
Then the total degree of polymerization $N$ is defined as $N \equiv N_A+N_B$.
We introduce an index $s$ to specify each segment, where $s=0$ corresponds to the free end of the $A$-block and $s=N$ corresponds to the other free end of the $B$-block.  Therefore, $0 \le s \le N_A$ and $N_A \le s \le N$ correspond to the $A$- block and the $B$-block, respectively.
In order to evaluate the conformational entropy, we need the statistical weight of any subchains.  
Let us use the notation $Q(s', {\bf r}'; s, {\bf r})$ to denote the statistical weight of a subchain between $s$-th and $s'$-th segments ($0 \le s' \le s \le N$) that are fixed at the positions ${\bf r}$ and ${\bf r}'$.  This statistical weight can be obtained by solving the following Edwards equation within the mean-field approximation 
\begin{equation}
  \frac{\partial}{\partial s } Q(s', {\bf r}'; s, {\bf r})
     = \Bigl[ \frac{b(s)^2}{6} \nabla^2 - \beta V(s, {\bf r}) \Bigr]
       Q(s', {\bf r}'; s, {\bf r}),
\label{Schroedinger equation1}
\end{equation}
where $\beta = 1/(k_B T)$, $b(s) = b_K$ if the $s$-th segment is the $K$-type segment, and $V(s, {\bf r})$ is an external potential acting on the $s$-th segment at ${\bf r}$ imposed by the surrounding segments.  Here, we assume that the external potential $V(s, {\bf r})$ is the same if the segment species ($A$ or $B$) is the same. 
 Thus,
\begin{eqnarray}
  V(s, {\bf r}) =
\left\{
	\begin{array}{rl}
	V_A({\bf r}) & \mbox{if $s$ indicates an $A$-segment} \\
	V_B({\bf r}) & \mbox{if $s$ indicates a  $B$-segment}. \\
	\end{array}
\right.
\end{eqnarray}
Equation (\ref{Schroedinger equation1}) should be supplemented by the initial condition $Q(0, {\bf r}'; 0, {\bf r})= \delta( {\bf r}' - {\bf r} )$. 
As the two ends of the block copolymer are not equivalent, we should introduce another statistical weight $\widetilde{Q}(s', {\bf r}'; s, {\bf r})$, which is calculated in the opposite direction along the chain starting from the free end $s=N$.

To reduce the computational cost, we define an integrated statistical weights $q(s, {\bf r})$ and $\widetilde{q}(s, {\bf r})$as follows:
\begin{eqnarray}
  q(s, {\bf r}) \equiv \int d{\bf r}' Q(0, {\bf r}'; s, {\bf r})  
  \nonumber \\
  \widetilde{q}(s, {\bf r}) \equiv \int d{\bf r}' \widetilde{Q}(0, {\bf r}'; s, {\bf r}). 
  \label{normal path integral of subchain}
\end{eqnarray}
It is easy to confirm that $q(s, {\bf r})$ and $\widetilde{q}(s, {\bf r})$ also satisfy eq.~(\ref{Schroedinger equation1}).

By using eqs. (\ref{normal path integral of subchain}), the density of the  $K$-type segments at position ${\bf r}$ is given by 
\begin{eqnarray}
  \phi_K({\bf r})
      = C  \int_{s \in K{\rm -block}} ds \ q(s, {\bf r}) \widetilde{q}(N-s, {\bf r}),
	\label{concentration field of polymer}
\end{eqnarray}
where $C$ is the normalization constant:
\begin{eqnarray}
      C =  \frac{M}{ \int d{\bf r}\ \int ds q(s, {\bf r}) \widetilde{q}(N-s, {\bf r}) } = \frac{M} { {\cal Z} }.
\end{eqnarray}
The parameter $M$ is the total number of chains in the system and ${\cal Z}$ is the single chain partition function which is independent of $K$, i.e. $ {\cal Z}=\int d{\bf r}q(s, {\bf r}) \widetilde{q} (N-s, {\bf r})= \int d{\bf r} q(N, {\bf r}) = \int d{\bf r} \widetilde{q}(N, {\bf r})$ .

The external potential $V_K({\bf r})$ can be decomposed into two terms as follows
\begin{eqnarray}
      V_K({\bf r}) =  \sum_{K'} \epsilon_{KK'} \phi_{K'}({\bf r}) - \mu_{K}({\bf r}).
      \label{external potential}
\end{eqnarray}
The first term is the interaction energy between segments and the $\epsilon_{KK'}$ is the nearest-neighbor pair interaction energy between a $K$-type segment and a $K'$-type segment, which is related to the Flory-Huggins interaction parameter via
$\chi_{AB} \equiv z \beta \bigl[
                             \epsilon_{AB}
                             - (1/2) (\epsilon_{AA} + \epsilon_{BB})
                       \bigr]
$
where $z$ is the number of nearest neighbor sites.
The $\mu_{K}({\bf r})$ is the chemical potential of the $K$-type segment, which is the Lagrange multiplier that fixes the density of the $K$-type segments at the position ${\bf r}$ to the specified density value.
The $V_{K}({\bf r})$ must be determined in a self-consistent manner so that this constraint is satisfied.
Such a self-consistent condition is achieved by an iterative refinement of the $V_{K}({\bf r})$. 

To improve the stability of the numerical scheme, we used the following finite difference scheme for the Edwards equation, eq. (\ref{Schroedinger equation1})
\begin{equation}
  q(s+ \Delta s, {\bf r}) 
     = \exp \bigl[
                  - \frac{\beta  V(s, {\bf r}) \Delta s } {2}
            \bigr]
            \Bigl(
                  1 + \frac{b(s)^2}{6} \nabla ^2 \Delta s
            \Bigr)
            \exp \bigl[
                  - \frac{\beta  V(s, {\bf r}) \Delta s } {2}
                 \bigr]
           q(s, {\bf r}).
\label{multistate eq}
\end{equation}

The Helmholtz free energy of the system can be given as follows
\begin{equation}
  {\cal F} = - k_{\rm B}T M \ln{{\cal Z}} + \frac{1}{2} \sum_{K} \sum_{K'} \int d{\bf r}\epsilon_{ KK'} \phi_K({\bf r}) \phi_{K'}({\bf r}) - \sum_K \int d{\bf r}V_K({\bf r}) \phi_K({\bf r}).
\label{free energy}
\end{equation}

To introduce dynamics into the model, we assume Fick's law of linear diffusion for the segment densities and an effect of the flow advection as follows
\begin{equation}
  \frac{\partial}{\partial t} \phi_K({\bf r},t) 
       = L_K \nabla^2 \mu_{K}({\bf r})
         - \nabla \{ {\bf v}({\bf r},t) \phi_K({\bf r},t) \},
\label{dynamical equation of diffusion}
\end{equation}
where $L_K$ is the mobility of $K$-type segment
and ${\bf v}({\bf r},t)$ is the local flow velocity such as the velocity of the externally imposed shear flow.

\subsection{System size optimization method}

Periodic microdomain structures of diblock copolymers have the crystal symmetry. 
To obtain equilibrium states of these periodic structures using the mean field theory, the free energy density of the system must be minimized with respect to the lattice structures of the ordered microdomains.
Same is true for two phase coexisting states where the system size should be optimized with respect to the coexisting two periodic structures.
For these purpose we introduce the system size optimization (SSO) method that minimizes the free energy density of the system by optimizing the side lengths of the simulation box on which periodic boundary conditions are imposed.
This is a similar method as the constant pressure molecular dynamics simulation proposed by 
Andersen\cite{Andersen}.
In the static SCF calculations, this optimization can be performed by requiring the following local equilibrium condition for each side length of the simulation box:
\begin{equation}
  \frac {\partial \cal F}{\partial \mathcal{L}_i} = 0, 
  \label{SSO static}
\end{equation}
where $\mathcal{L}_i(i=x,y,z)$ is the side length of the simulation box.
The left-hand side of eq. (\ref{SSO static}) can be evaluated numerically using the 
following central difference approximation
\begin{equation}
  \frac {\partial \cal F}{\partial \mathcal{L}_i} 
=  \frac { F (\mathcal{L}_i + \Delta \mathcal{L}_i ) 
- F ( \mathcal{L}_i - \Delta \mathcal{L}_i ) }
{ 2 \Delta \mathcal{L}_i },
  \label{SSO static numerial}
\end{equation}
where $\Delta \mathcal{L}_i$ is a small variation of $\mathcal{L}_i$.
We used the parabolic optimization method\cite{Numerical Recipe in C} to solve eq. (\ref{SSO static}).

On the other hand, when the dynamical SCF calculation is performed, we should regard $\mathcal{L}_i$ as a dynamical variable whose dynamics is described by the following ficticious equation of motion
\begin{equation}
  \frac {\partial \mathcal{L}_i} {\partial t} = - \zeta_i \frac {\partial \cal F}{\partial \mathcal{L}_i}, 
\label{dynamics_L}
\end{equation}
where $\zeta_i$ is a positive coefficient whose value is chosen properly so that the local equilibrium condition eq.(\ref{SSO static}) for $\mathcal{L}_i$ is guaranteed at every time step.

We checked the validity of our dynamical SSO method by using an A-B diblock copolymer melt whose stable equilibrium phase is the C phase. We performed two dimensional simulations where we assumed that $\zeta_x = \zeta_y = \zeta$ for simplicity, and we changed $\zeta$ from 0.0 to 0.5.
The parameters characterizing the A-B diblock copolymer are as follows:
the total length of the copolymer $N = 20$, the block ratio of the A block $f = N_A/N = 0.35$, and the effective bond lengths of each segment type are unity.  The interaction parameter is set to be $\chi N = 15$, which corresponds to the C phase in its equilibrium state
\cite{Matsen Schick}.
The initial state is set to the D phase to which we added small random noise with the standard deviation 0.0006. The initial shape of the simulation box is a square with side length $32.0$. As the square shape of the simulation box is not compatible with the perfect C phase, the SSO method adjusts the side lengths of the simulation box automatically.

In Figure 2, we show a comparison of the domain morphologies in the late stage ($t=5000$) between the two cases (a) with $\zeta = 0.001$ and (b) with $\zeta=0.05$, respectively.  In case (a), the C structure is distorted because the rate of the change in the side lengths of the simulation box is too slow to catch up with the change in the domain periodicity.  On the other hand in case (b), a perfect C phase is realized.
When $\zeta=0.5$, we observed that the dynamical scheme eq.(\ref{dynamics_L}) becomes unstable.

Other dynamical variables that depend on the value of $\zeta$ are shown in Figures 3 and 4.
Figure 3 shows the time evolution of the free energy. 
The dotted line is the reference state with $\zeta=0$ (i.e. the case without SSO) which reaches the distorted morphology shown in Figure 2(a). When the value of $\zeta$ is small ($\zeta=0$ and 0.001), it takes longer time for the free energy to relax and finally the system is trapped in a local minimum of the free energy.
For the intermediate values of $\zeta$ ($\zeta=0.05$, 0.1, and 0.2), the system reaches the perfect C phase as shown in Figure 2(b).  When the value of $\zeta$ is large ($\zeta=0.5$), the free energy initially drops rapidly and then the system is trapped by a local minimum of the free energy. These results mean that choosing an appropriate value of $\zeta$ accelerates the system to relax to the equilibrium domain morphology without distortions and defects.

Figure 4 shows the time evolution of the side lengths of the simulation box.  
The solid curves and the dotted curves indicate the $\mathcal{L}_x$ and the $\mathcal{L}_y$, respectively.
In all cases, the side lengths increase in the initial stage.  After such an initial stage, the side lengths reach their maximum values and then decrease for large $\zeta$ value.
When $\zeta = 0.05$, the curve does not show an overshoot, and the system smoothly reaches the perfect C phase. Thus, we judge that $\zeta=0.05$ is the most appropriate value for our system.

The above-mentioned dynamical SCF simulation can be performed with use of the "Simulation Utilities for Soft and Hard Interfaces (SUSHI)" in OCTA system
\cite{SUSHI}.  The simulation results reported in this article is obtained using SUSHI.

\section{Simulation Results} 

We simulated the epitaxial OOT G $\rightarrow$ C by imposing an external shear flow to an A-B diblock copolymer that is characterized by the parameters given in Section 2.2 using the technique described in the previous section.
The details of the simulation procedure are given below.

\subsection{Initial gyroid structure and final cylindrical structure}
The initial state of the simulation is chosen as the equilibrium G structure at $\chi N=20$.
To generate such an equilibrium G structure, we used the following procedure.
Let us denote the equilibrium (or steady state) side length of the unit cell of the G structure as $D_G$, and the equilibrium (steady state) spacing of the lamellar structure formed by the same block copolymer at $\chi N=20$ as $D_L$.  The value of $D_L$ can easily be obtained using a one dimensional SCF calculation with SSO.  
Then, assuming an epitaxial relationship in the transitions L\{10\} $\rightarrow$ C \{10\} $\rightarrow$ G \{211\} at a fixed value of $\chi N$, we can obtain an approximant for $D_G$ as follows
\begin{equation}
  D_G = \sqrt{ 6 } D_L.
  \label{D_G definition}
\label{G_SSO}
\end{equation}
Using this value of $D_G$ as the initial size of the simulation box, we set the SCF potential with the G symmetry as
\begin{equation}
  V(x,y,z) = V_0 \Bigl( \cos \frac{2 \pi x}{D_G} \sin  \frac{2 \pi y}{D_G} + \cos  \frac{2 \pi y}{D_G} \sin  \frac{2 \pi z}{D_G} + \cos  \frac{2 \pi z}{D_G} \sin  \frac{2 \pi x}{D_G} \Bigr) ^2,
  \label{G_level_surface}
\end{equation}
where $x$, $y$, and $z$ are the Cartesian coordinates, and the $V_0$ is an arbitrary small coefficient which we assume to be 0.001 for the minor segments and -0.001 for the major segments.
The use of the squared form on the right-hand side of eq.(\ref{G_level_surface}) originates from the fact that the gyroid structure in block copolymer melt is formed by double networks each with the G symmetry.  By assigning different signs to the $V_0$'s for major and minor segments, we can let the minor phase to gather inside the gyroid network while the major phase becomes rich in the matrix region.  

Starting from the SCF potential given by eq.(\ref{G_level_surface}), we perform a three dimensional static SCF calculation with SSO, which gives the equilibrium G structure.
Figure 5 shows the optimized bicontinuous double gyroid structure obtained using the above method, where the parameter $\Delta s$ in eq. (\ref{multistate eq}) is taken as 0.2. Figure 5 shows the super cell composed of eight optimized conventional unit cells of the G structure.
The side length of the optimized G unit cell is $D_G^0 = 17.2$.

From this G super cell, we can extract another unit cell as shown in Figure 6(a) where $X$, $Y$ and $Z$ axes are chosen to be parallel to the $[1\bar{1}0]$, $[11\bar{2}]$, and $[111]$ directions, respectively.
The obtained unit cell in Figure 6(a) is the minimal periodic unit cell with the $Z$ axis oriented to the [111] direction of the G unit cell.  
The side lengths of the unit cell are $\sqrt{2} D_G^0$, $\sqrt{6} D_G^0$, and $(\sqrt{3}/2) D_G^0$, respectively, where the volume of the unit cell is three times larger than that of the cubic G unit cell.

Figure 7 shows the projections of the G structure onto three different directions.
The bicontinuously arranged rods are the domains composed of the minor A phase.
Figure 7(a) shows the projection along the [111] direction that is the same as the left-hand side picture in Figure 6(a).
Figures 7(b) and 7(c) show the projections along the $[1\bar{1}0]$ and the $[11\bar{2}]$ directions, respectively.
In Figure 7, we can see the edges of the G unit cell (tilted cube) drawn by dotted lines and the extracted unit cell (cuboid) drawn by solid lines.

The self-consistent field on the $X$-$Y$ plane in Figure 6(a) is used as the initial condition for the static two dimensional SCF calculation for the C structure at $\chi N=15$, where we assumed an epitaxial OOT G $\rightarrow$ C. 
The optimized two dimensional C structure is shown in Figure 6(b), where we used the same scale as in Figure 6(a) for a direct comparison.

The lengths of the vertical and horizontal axes of the two dimensional C structure shown in Figure 6(b) are 2.0\% and 3.2\% larger than those of the G structure shown in Figure 6(a), respectively. 
As the changes in the side lengths are rather minor, we expect an epitaxial transition for the G structure at $\chi N=20$ to the C structure at $\chi N =15$.
The direction of the \{10\} plane of the cylindrical domains in Figure 6(b) coincides with that of the cylindrical domains in Figure 1(b).
This result contradicts the standard explanation of the epitaxial transition G \{211\} $\rightarrow$ C \{10\} which was proposed in the previous experimental works and mean field calculations. Instead, we expect that the actual epitaxial transition should be G \{220\} $\rightarrow$ C \{10\} as shown in Figure 1(b).

\subsection{The epitaxial OOT from the G structure to the C structure}

The OOT G $\rightarrow$ C is induced by a sudden increase in the temperature from $\chi N=20$ to $\chi N=15$, the former and the latter corresponding to the G and C phases, respectively
\cite{Matsen Schick}.
This phase transition is believed to be first order and should basically be driven by the thermal fluctuations.  
An introduction of an external flow accelerates the transition
\cite{Zvelindovsky}.
We introduce a shear flow whose direction is oriented to the [111] direction of the G unit cell.  
The velocity field ${\bf v}({\bf r})$ of this external shear flow is given by 
\begin{equation}
  {\bf v}({\bf r}) =  \Bigl( 0, 0, \dot{\gamma} (\frac{\mathcal{L}_y}{2} - y) \Bigr), 
\end{equation}
where $y$ is the Cartesian coordinate along the $Y$ axis and the $ \mathcal{L}_y /2$ is the $Y$-coordinate of the center of the system.
This flow field is indicated in Figure 6(a) by the arrows.
The Lees-Edwards boundary condition was employed in the $Y$ direction
\cite{Computer Simulation of Liquids}, and the periodic boundary conditions were employed in the other directions.

For the dynamical SCF calculation, the parameters were set as follows:
the criterion of the convergence of the segment density is $\Delta \phi = 0.0005$, i.e. if the difference between the two segment density fields at consecutive steps in the SCF iteration becomes everywhere below $\Delta \phi$, we regard the segment density field has converged.
The mobility $L_K$ in eq. (\ref{dynamical equation of diffusion}) is set to $L_K = 1.0$, the shear rate $\dot{ \gamma } = 0.001$, and $\Delta t = 0.01$, respectively.
The parameter $\zeta_i$ for the SSO is set 0.05 with which the SSO can reproduce the complete C domain in the two dimensional system as described in Section 2.2.  With this parameter, the SSO is performed at every other 100 time steps.

The temporal change of the microphase structure is shown in Figure 8. 
The G structure is deformed by the shear flow as shown in Figures 8(a) and (b). 
Suddenly, a grain boundary is generated in Figure 8(c), which is indicated by a white arrow.  This grain boundary consists of several cylinders parallel to the [111] direction of the G unit cell and this boundary region separates the upper G phase and the lower G phase.
The transition from the G structure to the C structure takes place in this lower phase as shown in Figures 8(d)-(f). 
The cylinders are tilted to the [111] direction of the G unit cell as shown in the side view of Figure 8(f).

The tilting of the cylinders is caused by the constant shear flow because a steady shear flow is composed of two contributions, a uniaxial extension and a rotation
\cite{kawakatsu book}.
This rotational contribution tilts the cylinders.
Such a tilting is suppressed in experiments by using oscillatory shear flows.

Three fold junctions in the upper G phase shown in Figure 8 (f) are stable and are migrated by the shear flow.  
Even in the late stage ($t$=10000), we cannot obtain the final equilibrium C structure.
This result suggests that boundary between the upper G phase and the lower C phase is stable and the separated G and C phases coexist stably.
Actually, a clear boundary between the G and C grains is observed by a polarized optical microscopy in a polymer solution
\cite{T. Q. Chastek T. P. Lodge}
and the long-lived coexistence between the G and the C phases is experimentally observed in a block copolymer 
\cite {Floudas Ulrich Wiesner Chu}.

The changes in the side lengths of the simulation box are shown in Figure 9.  
The side lengths in the $X$ and $Z$ directions are almost constant, which means that the epitaxial condition is satisfied. On the other hand, the side length in the $Y$ direction increases with time. 
The reason of the increase is explained below.

To check the effect of the SSO method, we carried out the same dynamical SCF simulation but without the SSO method.
The time evolution of the domain morphology is shown in Figure 10. 
In this case, as shown in Figures 10(a)-(d), the OOT occurs at the center of the system where the G structure transforms into the C structure.
The stable grain boundary as shown in Figure 8 is, however, not observed in this case without SSO.  The cylindrical domains are also tilted to the [111] direction of the G unit cell as is shown in the side view of Figure 10(c).
After such a transient state, the system reaches the complete C structure.
A characteristic phenomenon is observed near the center of the system where the cylindrical domains reconnect as shown in Figures 8(d)-(f).
Such reconnections continue steadily for certain time duration. 
This reconnection phenomenon means that the system is in a dynamical steady state where the energy injected by the shear flow into the system is released by the energy dissipation accompanied by the periodic reconnections of the cylindrical domains.

In order to check the stability of the systems, we show in Figure 11 the time evolution of the free energy density during the phase transition.
The free energy density in the case with the SSO shows a moderate change compared to that in the case without the SSO, which shows an oscillation synchronized to the reconnections of the cylindrical domains.  Such a periodic change in the free energy density is observed after $t=3000$ when the system reaches the almost perfect C structure without defects. In order to obtain such a perfect C phase, the system goes over energy barriers by the driving force of the shear flow.  In the case without SSO, however, the energy barrier should be much higher than the case with SSO because the condition of the constant system size imposes a sever restrictions on the reconnection of the cylindrical domains.  In the case with SSO, such a restriction is avoided by the increase in the side length of the simulation box in the $Y$ direction as shown in Figure 9.

We also tried simulations under a shear flow whose velocity gradient is set parallel to the $X$ direction. In this case, the free energy of the system increases slightly but the nucleation and growth of cylindrical domains can not be observed even in the late stage $t$=6000 either with SSO or without SSO. The epitaxial condition for the OOT is expected to be more precise for this direction of the velocity gradient than the case with the velocity gradient in the $Y$-direction. 
This is because the periodicity of the G structure in $X$ direction matches the periodicity of the C structure than that in the $Y$ direction, the former promoting the generation of the C structure. Contrary to this expectation, three fold junctions perpendicularly oriented to the [111] direction of the G unit cell continue to disconnect and reconnect due to the shear flow. 
Figure 12 shows this phenomenon. The circles in the Figure 12 indicate the three fold junctions where the disconnections and the reconnections take place. Figure 12(a) shows the structure after the disconnections, where we can observe the remains of three fold junctions indicated by the circles. Figure 12(b) shows the structure after the reconnections, where the three fold junctions regenerated. 
This result indicates that the G structure has different stabilities to different directions of the shear velocity gradient.

The reason of this different stability is explained by using Figure 13. Figure 13 shows three fold junctions perpendicularly oriented to the [111] direction of the G unit cell with different rotational angles to the direction of the shear gradient.
Figure 13(a) shows a three fold junction under a shear flow with its velocity gradient in the $Y$ direction.
The three domains extending from the center of the three fold junction are subjected to different shear flow velocities, i.e. the three domains do not move with the same velocity $v_y$. Thus, the three domains are elongated to the different directions with different rates, and the elongation finally makes the three fold junction disconnected. On the other hand, in the case of a three fold junction under a shear flow with its velocity gradient in the $X$ direction as shown in Figure 13(b), two domains have the same velocity $v_x$. In this situation, the three fold junction is elongated to the positive direction of the $X$-axis. Even after the disconnection, the separated domains keep closer and will be reconnected easily to form the three fold junction structure as shown in Figure 12.

\section{Discussions}

Our simulation showed that the epitaxial OOT G $\rightarrow$ C takes place in the [111] direction of the G unit cell and that the epitaxial relation for the G \{220\} $\rightarrow$ C \{10\} transition is achieved. The transition does not occur uniformly as shown in Figures 8(d)-(f) and Figure 10(c), where the C domain nucleates and grows.

Most of the experiments have reported the epitaxial OOT C \{10\} $\leftrightarrow$ G \{211\} which disagrees with our simulation result. A possible reason of this discrepancy is as follows. The first diffraction peak from a G structure is the peak from \{211\} and the intensity of this peak is stronger than that of \{220\} peak (secondary peak). 
Moreover, the positions of the peaks from G \{220\}, G \{211\} and C \{10\} are so close that it is not easy to judge which of the peaks from G \{220\} and G \{211\} epitaxially matches with the peak from C \{10\}. We calculated a three dimensional scattering function of the optimized G structure obtained in the simulation, and confirmed that the \{211\} spots are dominant and their intensities are about four times larger than those of the \{220\} spots.

If the G \{211\} $\rightarrow$ C \{10\} is realized under a shear flow with the velocity gradient in the $Y$ direction, the planes composed of cylinders are directed in parallel to the sheared plane, i.e. the $XZ$ plane. Thus, the friction generated by the reconnecting domains to the shear flow is expected to be smaller than that for the G \{220\} $\rightarrow$ C \{10\} case where the cylinder planes are perpendicular to the sheared plane. In our simulations, however, the system prefers the pathway as G \{220\} $\rightarrow$ C \{10\}. Therefore, we conclude that the direction of the velocity gradient is not an important factor in determining the direction of the C planes for the epitaxial transition. On the other hand, we confirmed that matching between the lattice constants is more important. As is shown in Figure 6(b), the origin of the selection of the generated C structure from the G [111] plane (Figure 6(a)) is the matching between the lattice constants. That is, the system prefers the kinetic pathway that minimizes the free energy of the system by matching the lattice constants, which leads to the G \{220\} $\rightarrow$ C \{10\} transition.

Previous theoretical studies have also supported the OOT G \{211\} $\rightarrow$ C \{10\}.
These studies relied on the reciprocal space representations.
However most of the experiments have been done under a condition with a shear flow and a temperature change.
We succeeded in reproducing such experimental conditions in our simulation. Using this simulation, we could reproduce the correct kinetic pathway of the epitaxial OOT, i.e. the nucleation and growth process of the C domains.

We found the difference in the stability of the G domains to the shear gradient direction due to the different velocities of the shear flow imposed on the three domains meeting at a three fold junction as shown in Figures 12 and 13.
The G structure is stable under the shear flow with the velocity gradient in the $X$ direction.
There has been no answer to the question why the complex G phase with three dimensional bicontinuous structure is generated under a shear flow. 
Our result demonstrates that the G structure is actually stable under a shear flow.

Although the detail of the transition process is complex, we understand that the three fold junctions with domains perpendicular to the [111] direction of the G unit cell do not play an important role in the transformation from G to C. 
Three fold junctions are simply disconnected and vanish during the phase transition.
This observation does not agree with the model of the epitaxial transition proposed by Matsen, where a three fold junction is connected to one of the nearest neighbor three fold junctions to form a five fold junction. In our observation, three fold junctions are stable and they are not connected to any other junctions.

Here, we propose a model of the kinetic pathway.
Figure 14(a) shows a projection of the G unit cell along the [111] direction, where the bold triangles are the projections of consecutive three domains.  
Figure 14(b) is the same structure as Figure 14(a) observed from a different direction.  
We can confirm that the triangles in Figure 14(a) are formed by consecutive three domains (shown in black) connected by three fold junctions.
When a shear flow is imposed, these black domains in Figure 14(b) are elongated and form cylinders as shown in Figure 14(c). 
These cylinders are rearranged to form a hexagonally packed cylindrical structure whose lattice spacings satisfy the epitaxial relations G \{220\} $\rightarrow$ C \{10\} as shown in Figure 14(d).
This model of the kinetic pathway can be verified in Figures 8(d)-(e) and in Figures 10(b)-(c).

\section{Conclusion}

The epitaxial OOT G $\rightarrow$ C was studied using the real space dynamical SCF technique with the SSO method. 
With such an SSO method, we succeeded in reproducing the realistic kinetic pathway of the first order phase transition of G $\rightarrow$ C.  On the other hand, in the absence of the SSO, we found that the kinetic pathway is very different from what we observe with SSO.
We also found that the G structure shows different responses to different directions of the velocity gradient of the shear flow.

Using this technique, we studied the kinetic pathway of the G $\rightarrow$ C transition induced by a shear flow in the [111] direction of the unit cell of the G structure.  
We observed the following kinetic pathway: 
The G domains perpendicularly oriented to the [111] direction of the G unit cell do not contribute to the formation of the cylindrical domains. They are disconnected and vanish during the transition.  
On the other hand, the other G domains are elongated by the shear flow and transform into the cylindrical domains. Such deformations occur locally, and the cylindrical domains are rearranged to form a hexagonally close-packed C structure.

The most important result of our simulations with SSO is that we can observe a nucleation and growth of the C phase in the matrix of the G phase, i.e. the nature of the first order phase transition, which was not observed in the previous simulations done in the Fourier space.  Under a steady shear flow, we observed that the G domains around the nucleus of the C phase deform and gradually join the C phase. 
We also observed that the domain spacing satisfies the epitaxial relationship G \{220\} $\rightarrow$ C \{10\} as was proposed in the experimental work
\cite{M. F. Schulz F. S. Bates K. Almdal K. Mortensen}. 
We could not observe the transformation process of the domains from three fold junctions to five fold junctions as was previously proposed
\cite{Matsen Gyroid}.

We found that the dynamical SCF theory with the SSO method in real space is very useful and reliable to trace the OOTs and ODTs between the microdomain structures of block copolymer melts.
\newline
\\
{\Large\bf Acknowledgment} \\

T. H. thanks Dr. H. Kodama and Dr. R. Hasegawa for the fruitful collaborations in coding SUSHI.
The authors thank Prof. M. Doi (Tokyo University) and the members of the OCTA project for many helpful comments and discussions.  
This study is executed under the national project, which has been entrusted to the Japan Chemical Innovation Institute by the New Energy and Industrial Technology Development Organization (NEDO) under METI's Program for the Scientific Technology Development for Industries that Creates New Industries.
This work is partially supported by Grant-in-Aid for Science from the Ministry of Education, Culture, Sports, Science and Technology, Japan. The computation was in part performed at the Super Computer Center of the Institute of Solid State Physics, University of Tokyo.

\newpage
{\Large\bf Figure captions}
\newline\newline
Figure 1.  A projection of the G unit cell structure observed from the $[111]$ direction
\cite{M. E. Vigild}. 
The circles indicate the positions of the cylindrical domains in the epitaxial transition (a) G $d_{\{211\}}$ $\rightarrow$ C $d_{\{10\}}$ and (b) G $d_{\{220\}}$ $\rightarrow$ C $d_{\{10\}}$, respectively.
\newline\newline
Figure 2.  A comparison of the domain morphologies of an A-B diblock copolymer obtained with the two dimensional dynamical SCF simulations with SSO. The simulations were started form a D phase without the external flow. 
The model parameters are shown in the text. The values of the SSO parameter $\zeta$ are (a) $\zeta=0.001$ and (b) $\zeta=0.05$, respectively.
\newline\newline
Figure 3.  The time evolutions of the free energy per chain for the C structure for several $\zeta$ values are shown. The dotted curve is the result of the reference simulation with $\zeta=0$.
\newline\newline
Figure 4. The time evolutions of the side lengths of the simulation box for several $\zeta$ values are shown.  The solid and dotted curves show $\mathcal{L}_x$ and $\mathcal{L}_y$, respectively.  Both side lengths are normalized using their initial values $\mathcal{L}_{i_0}=32$ as $\mathcal{L}_i / \mathcal{L}_{i_0}$. 
\newline\newline
Figure 5. The isosurfaces of the super cells ($2 \times 2 \times 2$) of a bicontinous double gyroid structure for $\chi N=20$, $\phi=0.9$ are shown.  The single unit cell is calculated using $32 \times 32 \times 32$ meshes.
\newline\newline
Figure 6. A comparison of the G structure and the two dimensional C structure obtained by the static SCF simulations with SSO is shown.  These two structures indicate an epitaxial OOT G $\rightarrow$ C in the [111] direction.  
(a) A view of the isosurfaces of the minimal periodic structure of the G unit cell along the [111] direction is shown, The parameter are chosen as $\chi N = 20$, and $\phi = 0.75$ (left-hand side). The velocity distribution of the added shear flow in the [111] direction (right-hand side). (b) The two dimensional C structure obtained as an equilibrium state starting from the domain structure on the cross section on a \{111\} plane of Figure 6 (a) with $\chi N = 15$. 
\newline\newline
Figure 7. Projections of the real space G structure onto different directions are shown, where the bicontinuously arranged rods correspond to the minor A phase of the G structure. Shown are the projections along (a) the [111] direction, (b) the $[1\bar{1}0]$ direction, and (c) the $[11\bar{2}]$ direction, respectively. In (a), two different unit cells are shown: one is the conventional cubic unit cell drawn by dotted lines and the other is the parallelepiped unit cell which has a periodicity in the [111] direction drawn in solid lines.   The relation between the side lengths of these two unit cells are discussed in the text.
\newline\newline
Figure 8. The time evolution of the domain in the epitaxial OOT G $\rightarrow$ C simulated with SSO is shown. A shear flow is imposed in the $[111]$ direction of the G unit cell with $\dot{ \gamma } = 0.01$.  The directions of the shear flow are shown by black arrows.  The view point of the graphics is set so that one can verify the growth of the C structure. The snapshot figures are taken at times $t$ = (a) 100, (b) 960, (c) 1960, (d) 3960, (e) 4960, and (f) 5960, respectively.
For the cases (c) and (f), we also show the side views, where the white arrow indicates the stable grain boundary where cylindrical domains parallel to the shear direction ($[111]$ direction) is generated.
\newline\newline
Figure 9. The time evolutions of the side lengths of the simulation box during the phase transition are shown.
\newline\newline
Figure 10. Similar to Figure 8 but without SSO.  
\newline\newline
Figure 11. The time evolutions of the free energy per polymer chain with SSO and without SSO during the phase transition are shown. In both of these cases, the initial value of the free energy is set to be zero.
\newline\newline
Figure 12. The time evolution of the G domain under a shear flow with its velocity gradient in the $X$ direction and with $\dot{ \gamma } = 0.01$. The circles indicate the three fold junction perpendicularly oriented to the [111] direction of the G structure. (a) Three fold junctions are disconnected at $t$=3600. (b) Three fold junctions are regenerated at $t$=4200.
\newline\newline
Figure 13. Three fold junctions are migrated to different directions by the shear flow. (a) Under a shear flow with velocity gradient in the $Y$-direction and (b) a shear flow with velocity gradient in the $X$-direction. In these two cases, the domains migrate in different manners.
\newline\newline
Figure 14. The model of the kinetic pathway of the epitaxial OOT G $\rightarrow$ C is illustrated.  (a) Shown is the view of the G unit cell along the [111] direction, where the black domains transform to the C domains. (b) Another view of the same G unit cell as in (a) from a different direction is shown. (c) An image of the transformation from the elongated domains to the cylinders is shown. (d) Positions of the rearranged cylinders with hexagonal order are shown, where the lattice spacings satisfy the epitaxial relations G $d_{\{220\}}$ $\rightarrow$ C $d_{\{10\}}$.

\newpage
\begin{figure}[H] \includegraphics[width=0.55\linewidth]{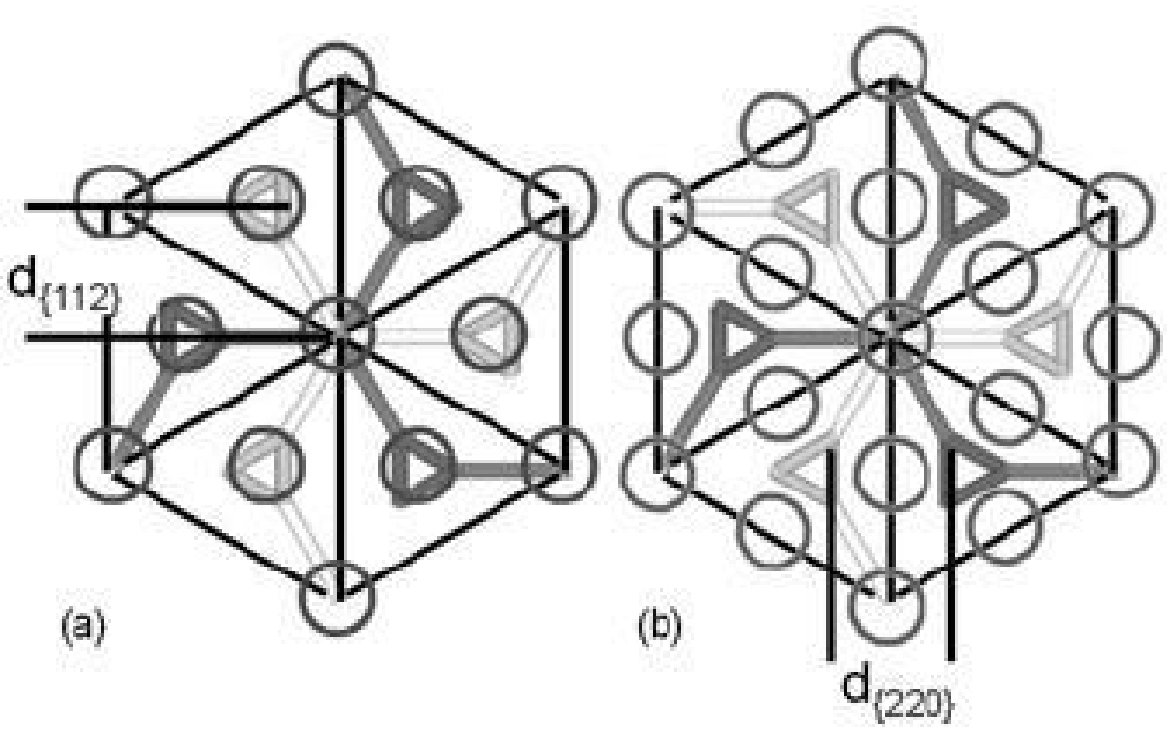} \caption{ } \end{figure}
\begin{figure}[H] \includegraphics[width=0.55\linewidth]{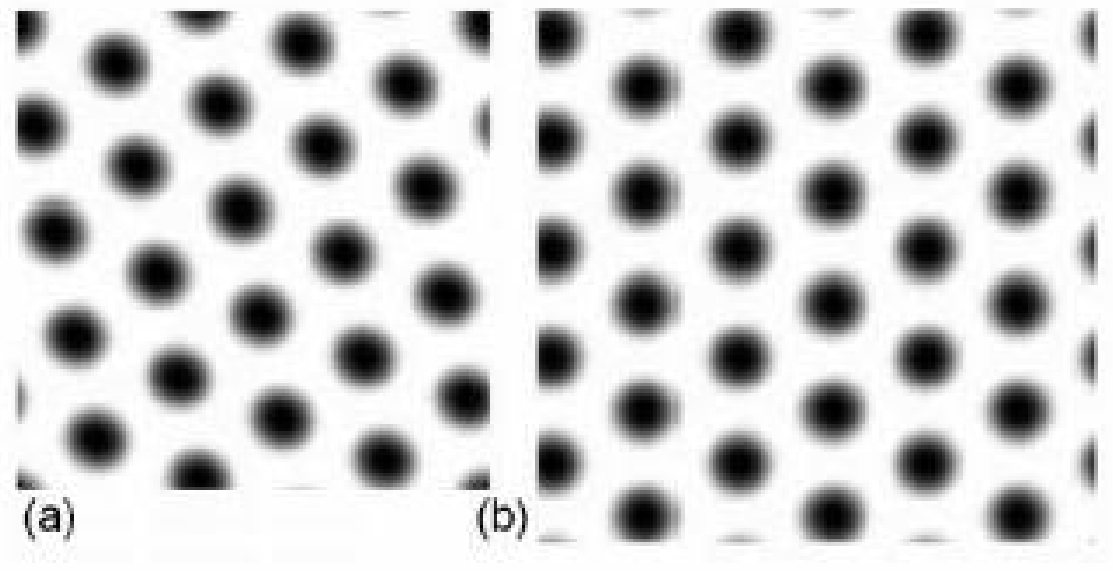} \caption{ } \end{figure}
\begin{figure}[H] \includegraphics[width=0.55\linewidth]{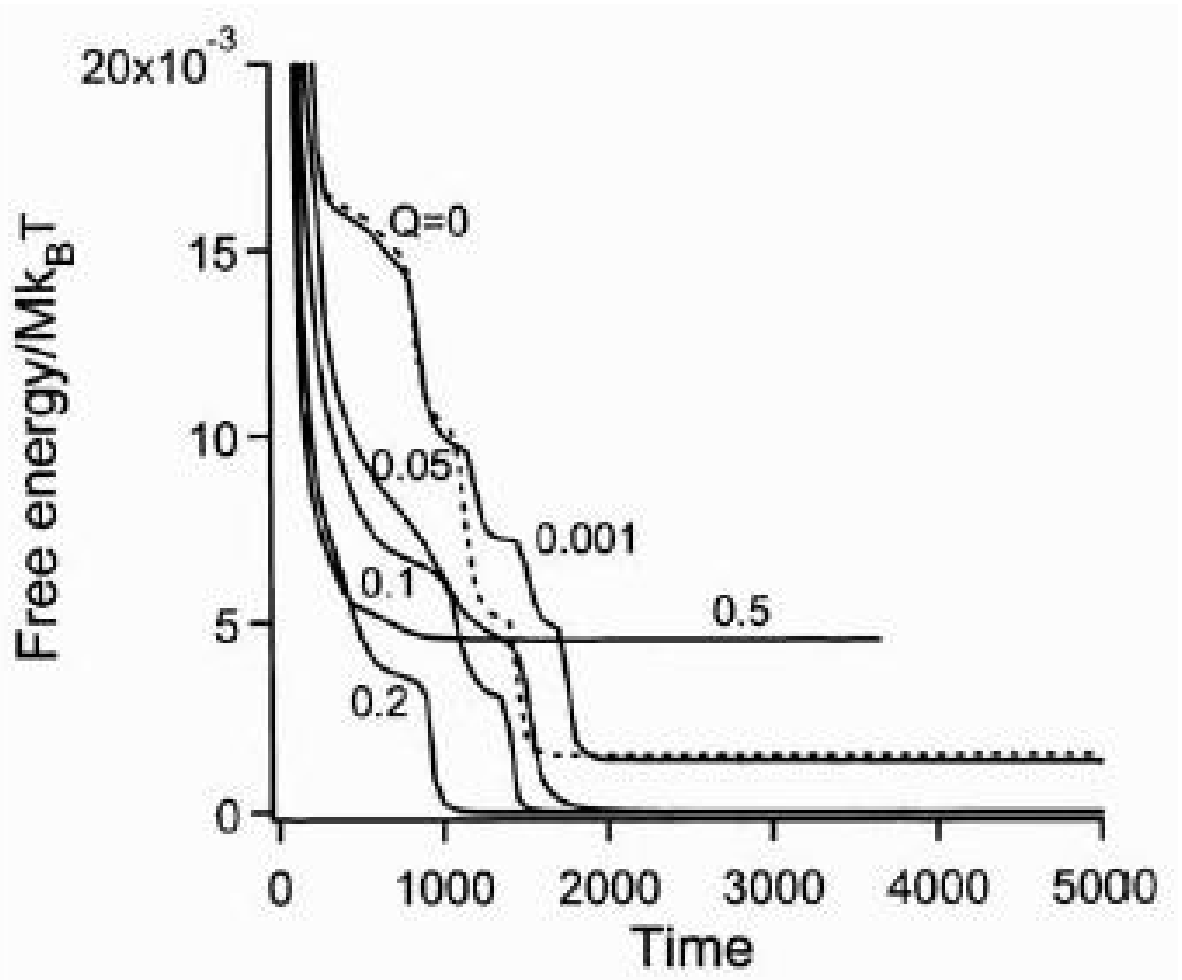} \caption{ } \end{figure}
\begin{figure}[H] \includegraphics[width=0.55\linewidth]{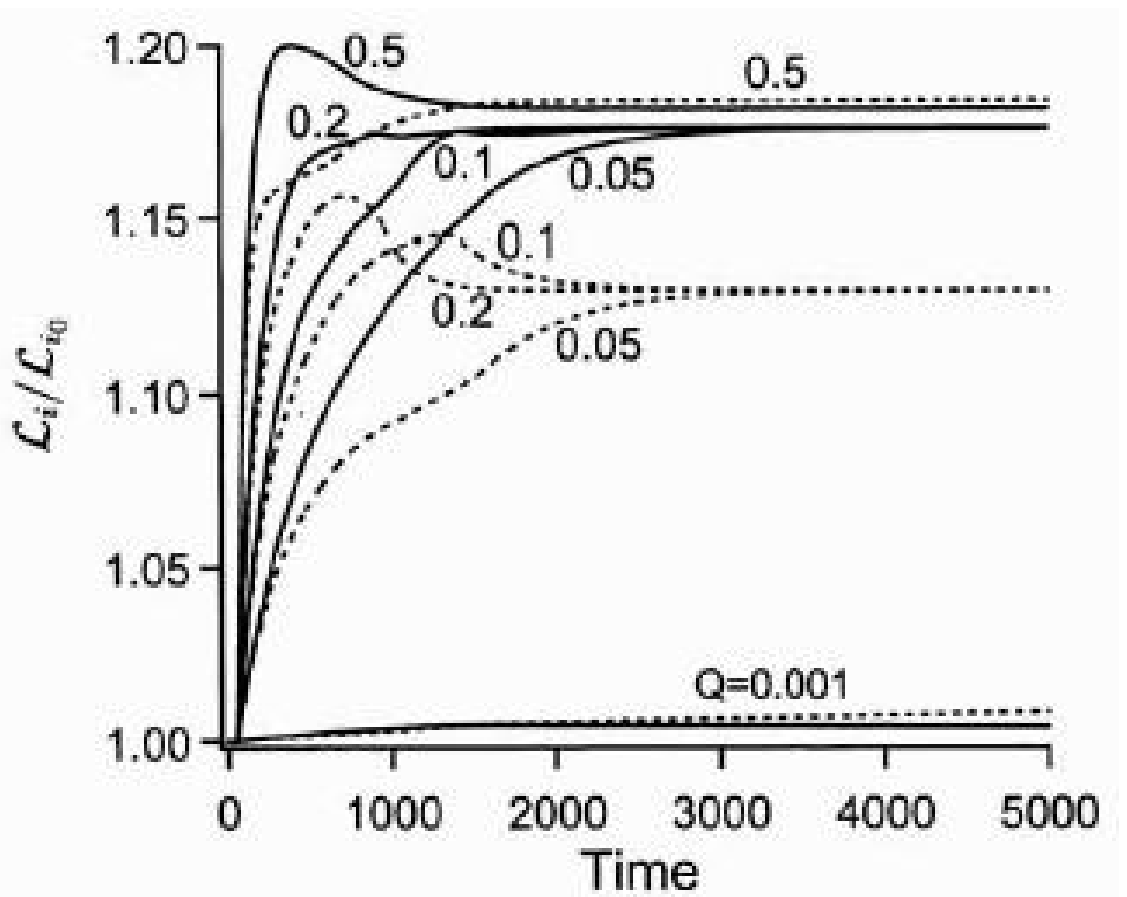} \caption{ } \end{figure}
\begin{figure}[H] \includegraphics[width=0.55\linewidth]{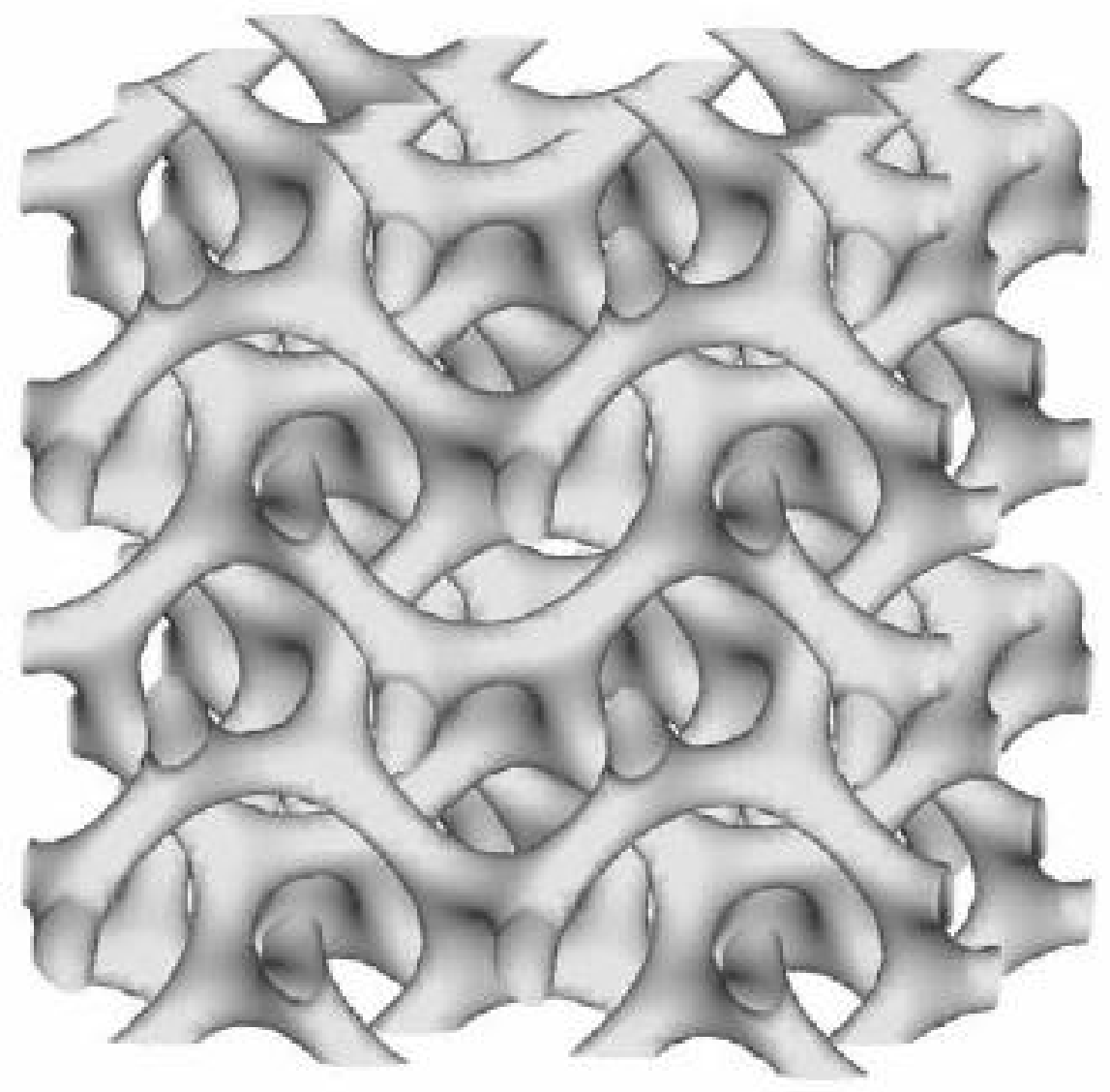} \caption{ } \end{figure}
\begin{figure}[H] \includegraphics[width=0.55\linewidth]{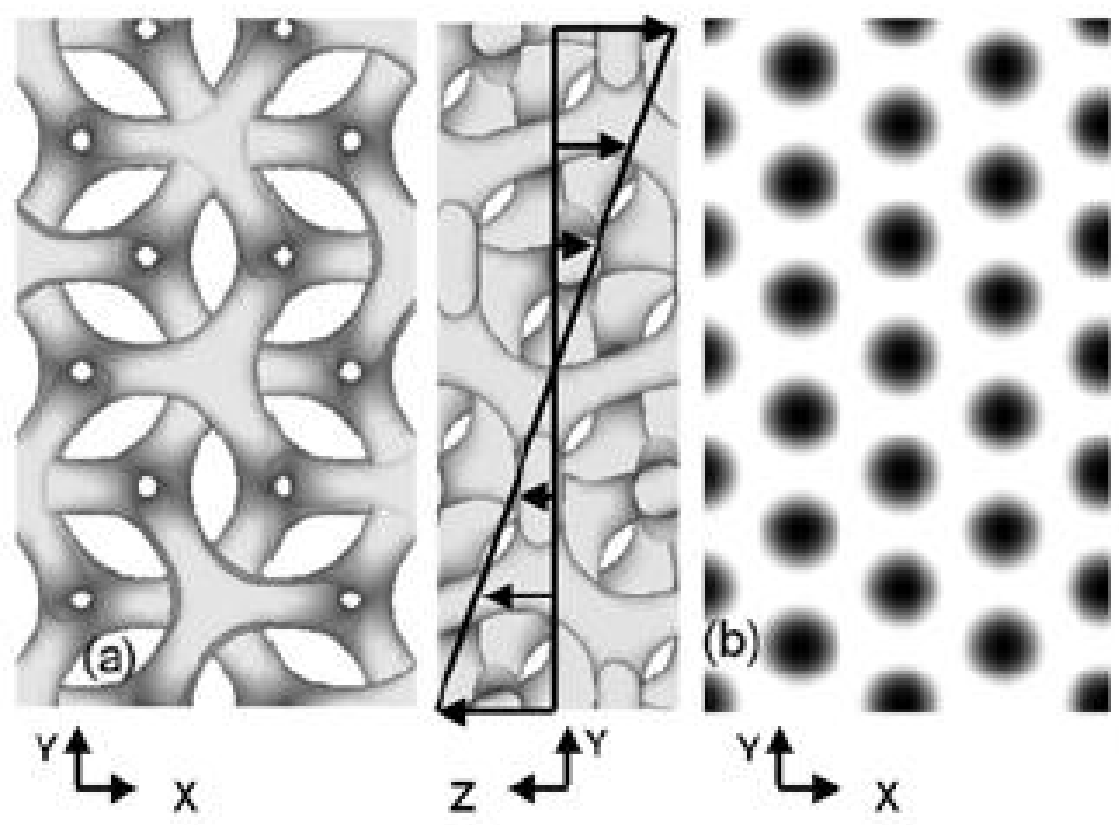} \caption{ } \end{figure}
\begin{figure}[H] \includegraphics[width=0.55\linewidth]{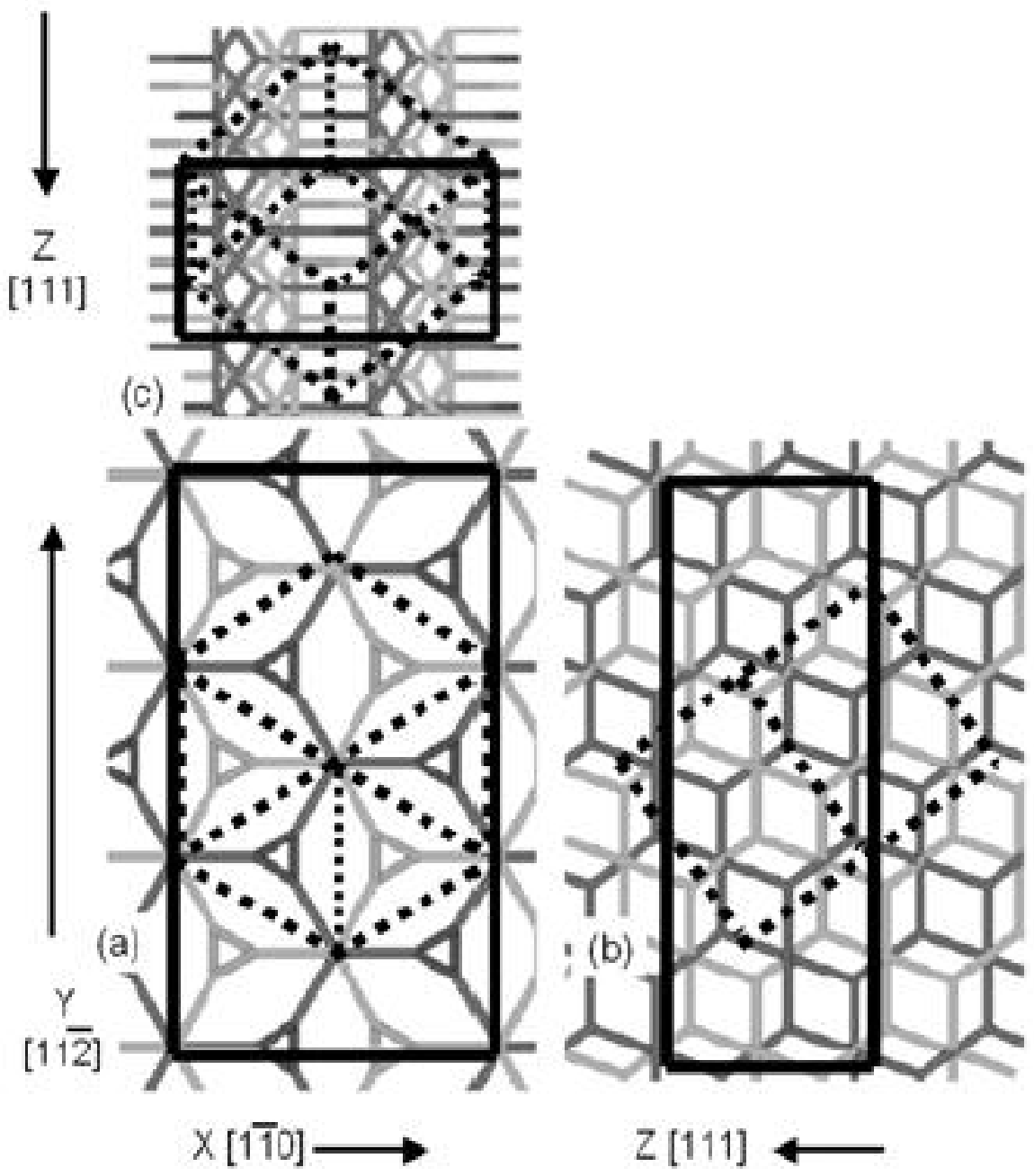} \caption{ } \end{figure}
\newpage 
\begin{figure}[H] \includegraphics[width=1.0\linewidth]{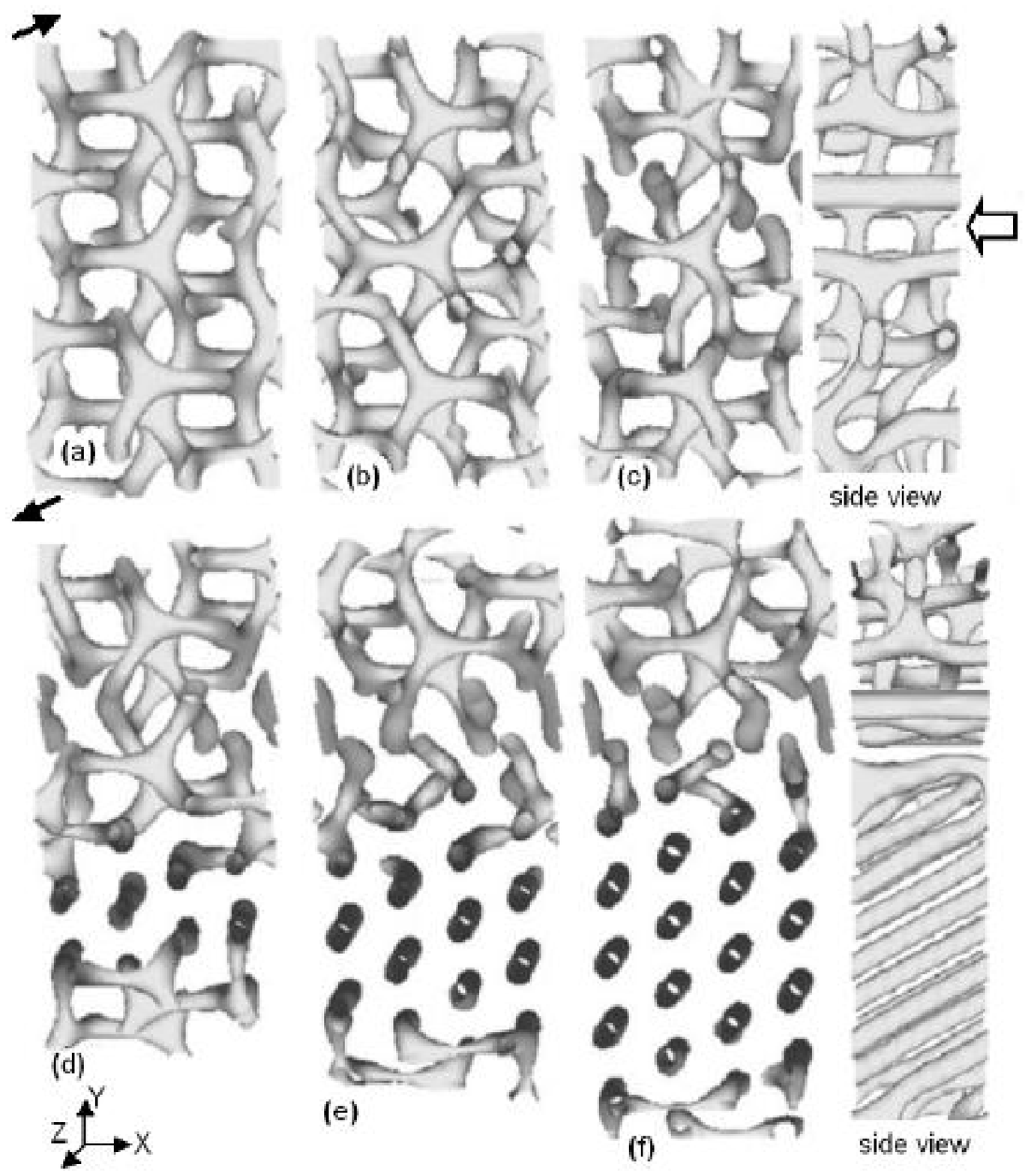} \caption{ } \end{figure}
\newpage 
\begin{figure}[H] \includegraphics[width=0.55\linewidth]{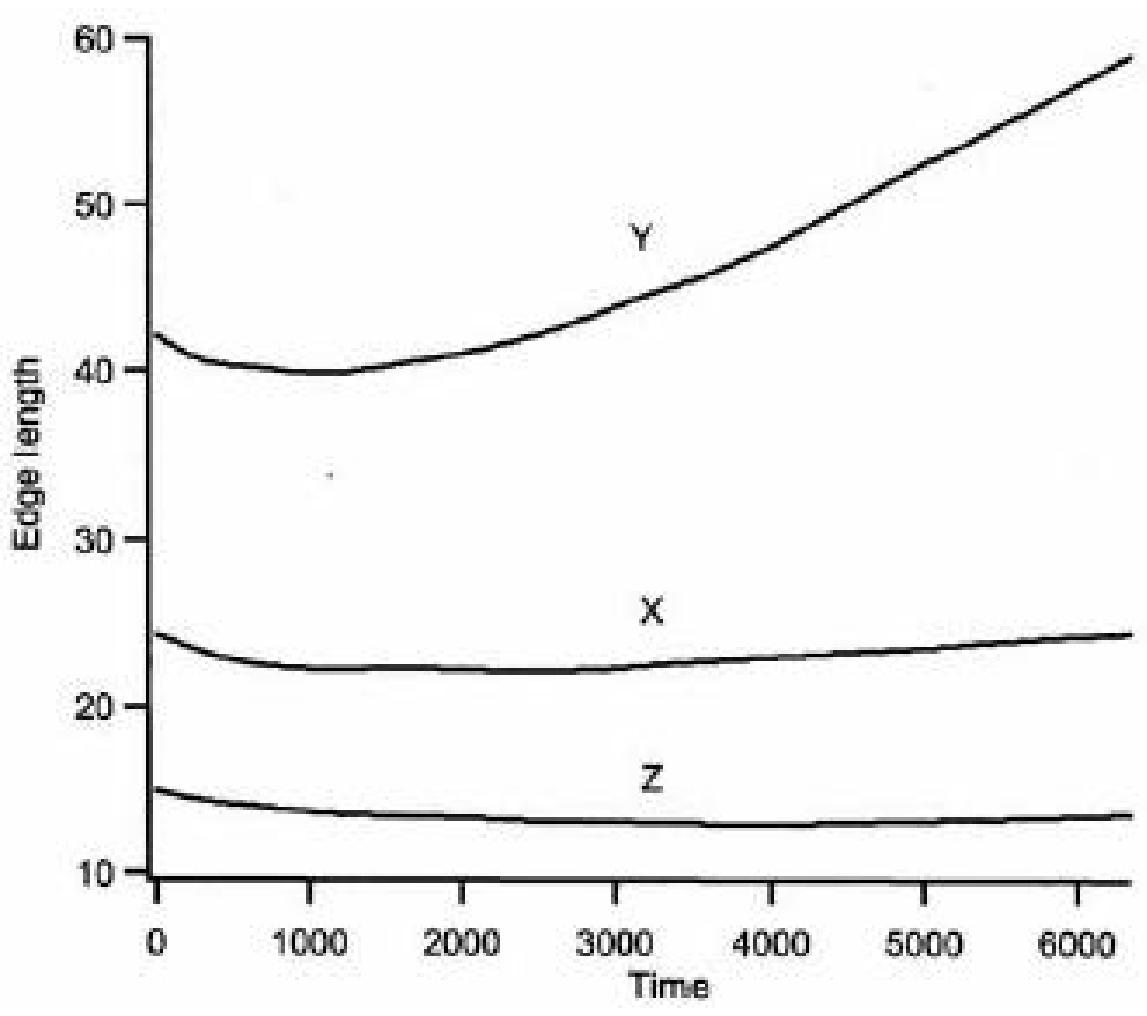} \caption{ } \end{figure}
\newpage 
\begin{figure}[H] \includegraphics[width=1.0\linewidth]{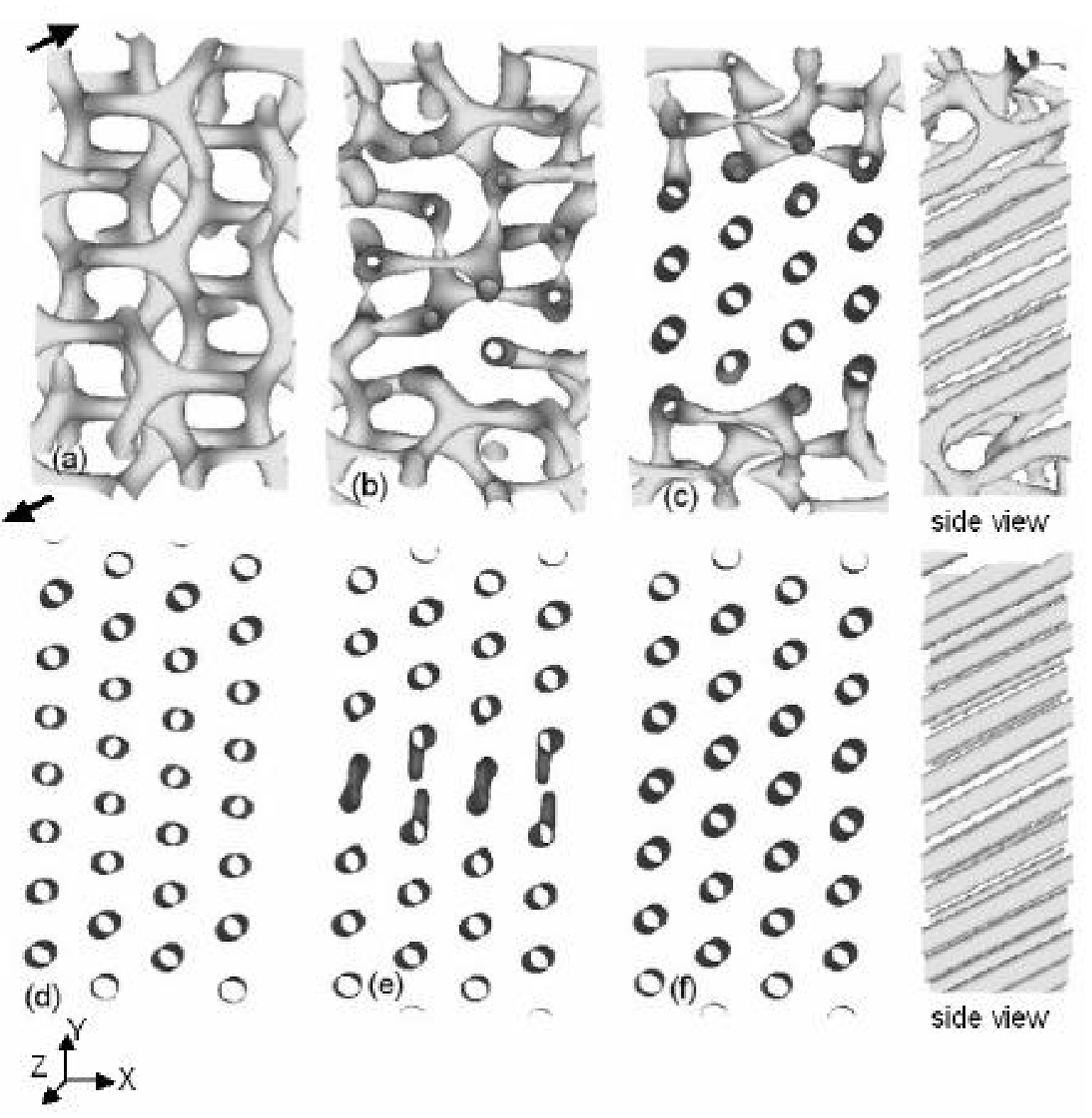} \caption{ } \end{figure}
\begin{figure}[H] \includegraphics[width=0.55\linewidth]{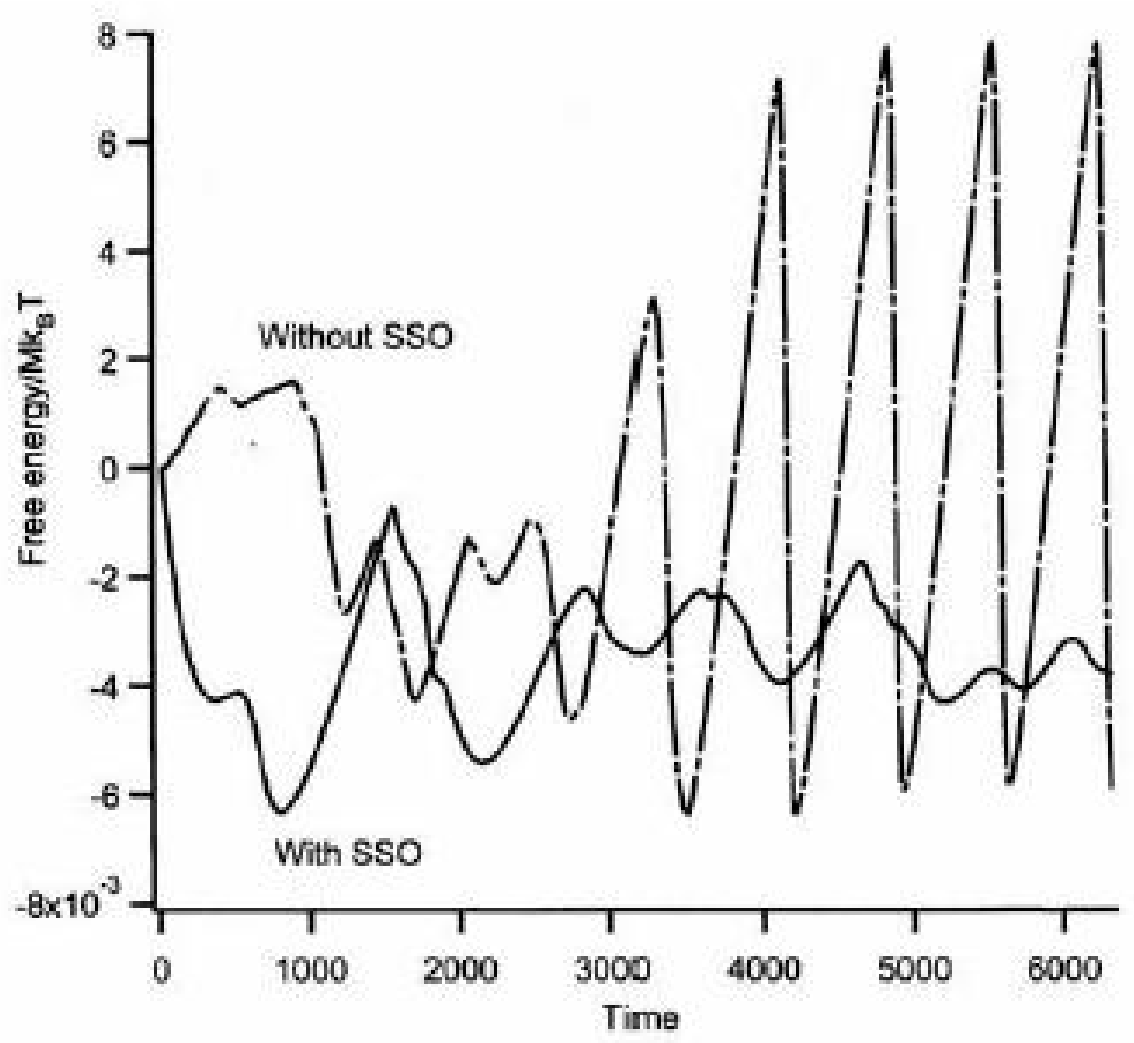} \caption{ } \end{figure}
\begin{figure}[H] \includegraphics[width=0.55\linewidth]{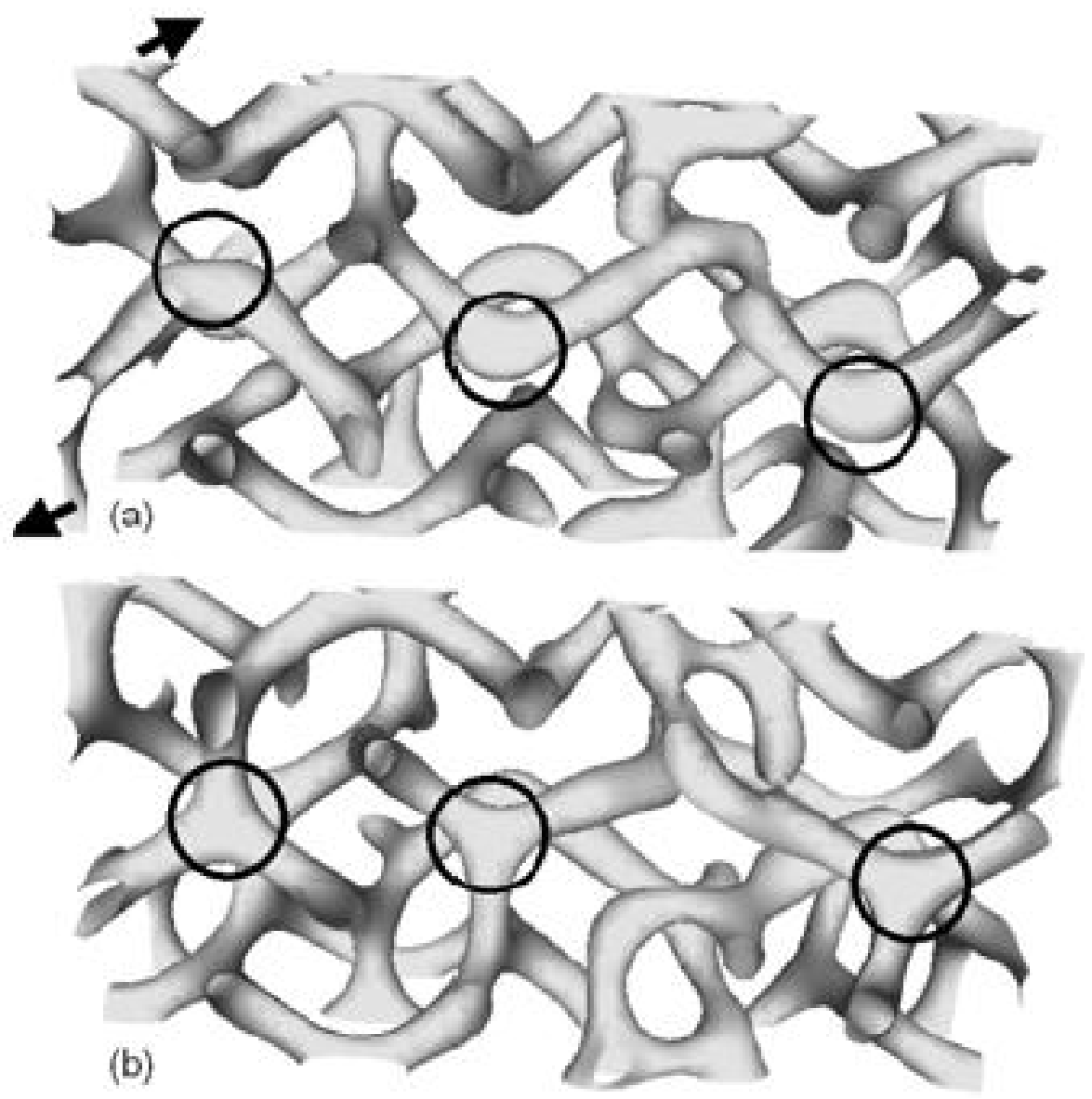} \caption{ } \end{figure}
\begin{figure}[H] \includegraphics[width=0.55\linewidth]{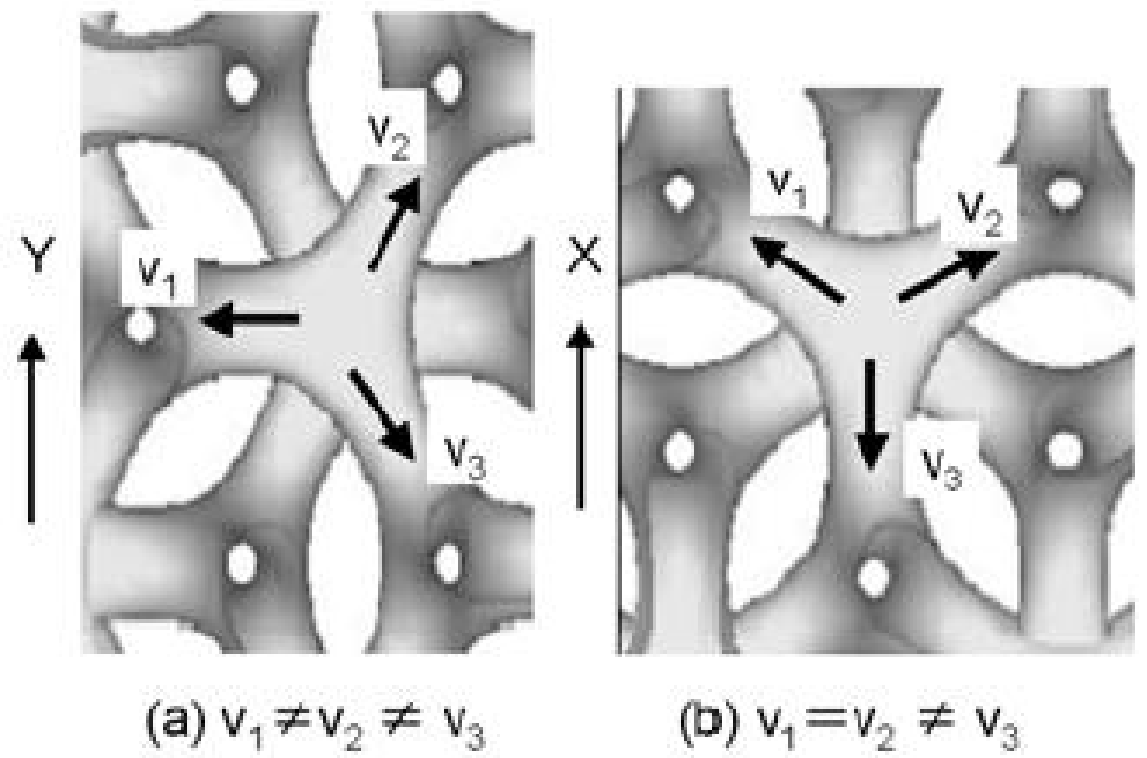} \caption{ } \end{figure}
\begin{figure}[H] \includegraphics[width=0.55\linewidth]{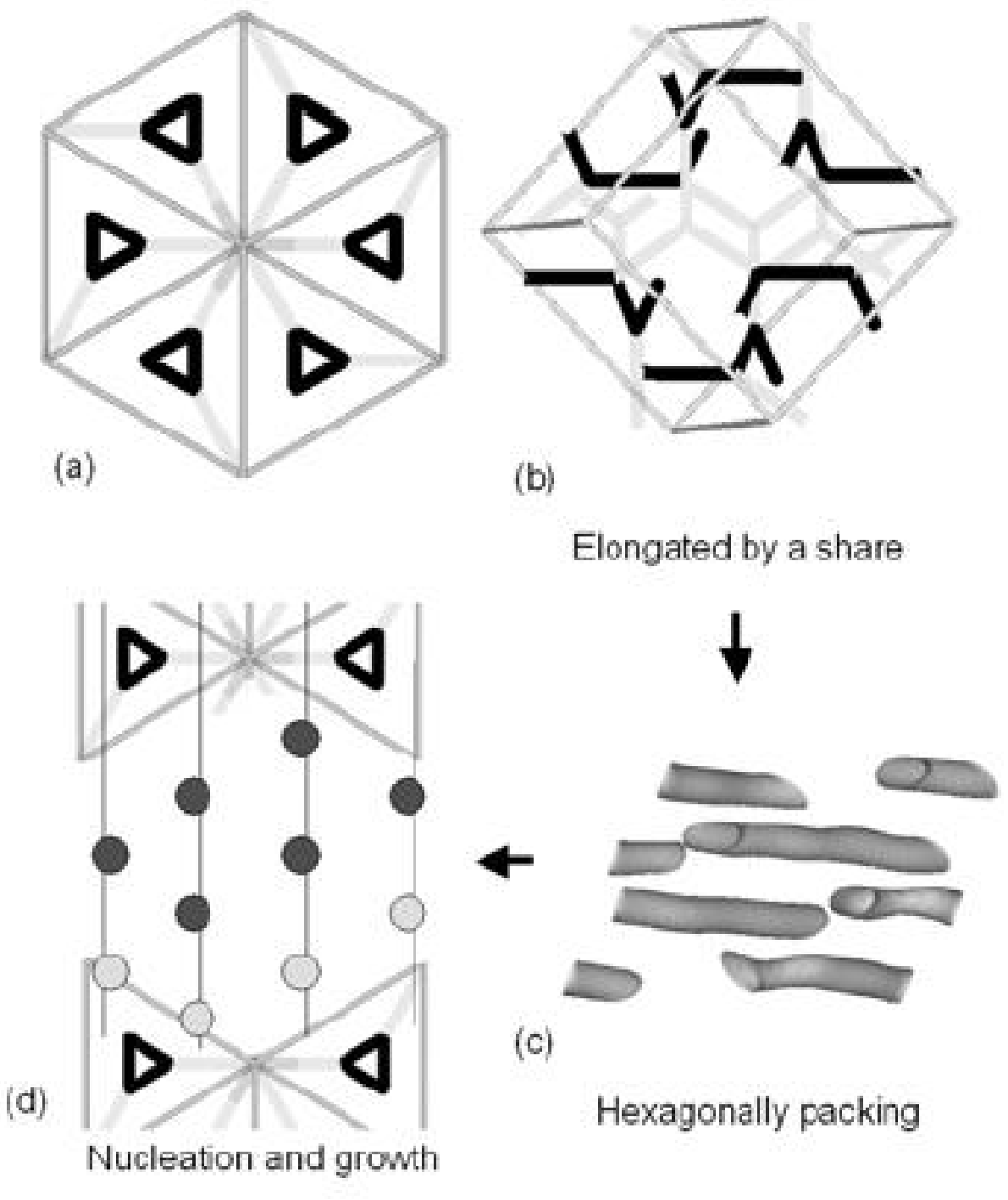} \caption{ } \end{figure}


\newpage
\begin{thebibliography}{45}

\bibitem{Matsen Bates} 
Matsen, M. W.; Bates, F. S.
\textit{Macromolecules} \textbf{1996}, 29, 1091-1098.

\bibitem{Hamley book}
Hamley, I. W.
\textit{Block Copolymers}; Oxford University Press: Oxford, 1999.

\bibitem{Bates Fredrickson} 
Bates, F. S.; Fredrickson, G. H.
\textit{Physics Today} \textbf{1999}, 52, 32-38.

\bibitem{Fredrickson}
Fredrickson, G. H.; Ganesan, V.; Drollet, F. 
\textit{Macromolecules} \textbf{2002}, 35, 16-39.

\bibitem{Hajduk}
Hajduk, D. A.; Harper, P. E.; Gruner, S. M.;
Honeker, C. C.; Kim, G.; Thomas, E. L.;
Fetters, L. J.
\textit{Macromolecules} \textbf{1994}, 27, 4063-4075.

\bibitem{Matsen Bates 96}
Matsen, M. W.; Bates, F. S.
\textit{J. Chem. Phys} \textbf{1997}, 106, 2436-2448.

\bibitem{Hashimoto}
Hashimoto, T.; Tsutsumi, K.; Funaki, Y.
\textit{Langmuir} \textbf{1997}, 13, 6869-6872.

\bibitem{Zhao}
Zhao, D.; Feng, J.; Huo, Q.; Melosh, N.; Fredrickson, G. H.; Chmelka, B. F.; Stucky, G. D.
\textit{Science} \textbf{1998}, 279, 548-552.

\bibitem{Chan}
Chan, V. Z.-H.; Hoffman, J.; Lee, V. Y.; Iatrou, H.; Avgeropoulos, A.; Hadjichristidis, N.; Miller, R. D.
\textit{Science} \textbf{1999}, 286, 1716-1719.



\bibitem{Y. Rancon J. Charvolin}
Ran\c{c}on, Y.; Charvolin, J.
\textit{J. Phys. Chem} \textbf{1988}, 92, 2646-2651.

\bibitem{M. F. Schulz F. S. Bates K. Almdal K. Mortensen}
Schulz, M. F.; Bates, F. S.; Almdal, K.; Mortensen, K.
\textit{Phys. Rev. Lett.} \textbf{1994}, 73, 86-89.

\bibitem{Forster}
F\"orster, S.; Khandpur, A. K.; Zhao, J.; Bates, F. S.; Hamley, I. W.; Ryan, A. J.; Bras, W.
\textit{Marcomolecules} \textbf{1994}, 27, 6922-6935.

\bibitem{M. E. Vigild}
Vigild, M. E.; Almdal, K.; Mortensen, K.; Hamley, I. W.; Fairclough, J. P. A.; Ryan, A. J.
\textit{Marcomolecules} \textbf{1998}, 31, 5702-5716.

\bibitem{Floudas Ulrich Wiesner Chu}
Floudas, G; Ulrich, R.; Wiesner, U.; Chu, B.
\textit{Europhys. Lett.} \textbf{2000}, 50, 182-188.

\bibitem{C. -Y. Wang T. P. Lodge}
Wang, C. Y.; Lodge, T. P.
\textit{Macromolecules} \textbf{2002}, 35, 6997-7006.

\bibitem{T. Q. Chastek T. P. Lodge}
Chastek, T. Q.; Lodge, T. P.
\textit{Macromolecules} \textbf{2003}, 36, 7672-7680.

\bibitem{Helfand Wasserman 4}
Helfand, E.; Wasserman, Z. R.
\textit{Macromolecules} \textbf{1976}, 9, 879-888.

\bibitem{Helfand Wasserman 5}
Helfand, E.; Wasserman, Z. R.
\textit{Macromolecules} \textbf{1978}, 11, 960-966.

\bibitem{Helfand Wasserman 6}
Helfand, E.; Wasserman, Z. R.
\textit{Macromolecules} \textbf{1980}, 11, 994-998.

\bibitem{Leibler}
Leibler, L.
\textit{Macromolecules} \textbf{1980}, 13, 1602-1617.

\bibitem{Matsen Schick}
Matsen, M. W.; Schick, M.
\textit{Phys. Rev. Lett.} \textbf{1994}, 72, 2660-2663.

\bibitem{Khandpur et al.}
Khandpur, A. K.; Foerster, S.; Bates, F. S.; Hamley, I. W.; Ryan, A. J.;Bras, W.;Almdal, K.; Mortensen, K. 
\textit{Macromolecules} \textbf{1995}, 28, 8796-8806.

\bibitem{Qi Wang 1}
Qi, S.; Wang, Z. G.
\textit{Phys. Rev. Lett.} \textbf{1996}, 76, 1679-1682.
\bibitem{Qi Wang 2}
Qi, S.; Wang, Z. G.
\textit{Pys. Rev. E} \textbf{1997}, 55, 1682-1697.
\bibitem{Qi Wang 3}
Qi, S.; Wang, Z. G.
\textit{Polymer} \textbf{1998}, 39, 4639-4648.

\bibitem{Nonomura Ohta 1}
Nonomura, M.; Ohta, T.
\textit{J. Phys. Soc. Jpn.} \textbf{2001}, 70, 927-930.
\bibitem{Nonomura Ohta 2}
Nonomura, M.; Ohta, T.
\textit{Physca A} \textbf{2002}, 304, 77-84.
\bibitem{Nonomura Ohta 3}
Nonomura, M.; Ohta, T.
\textit{J. Phys.: Condens. Matt.} \textbf{2003}, 15, L423-L430.

\bibitem{Hong}
Hong K. M.; Noolandi, J.
\textit{Macromolecules} \textbf{1981}, 14, 727-736.

\bibitem{Fleer}
Fleer, G. J.; Cohen Stuart, M. A.; Scheutjens, J. M. H. M.; Cosgrove, T.; Vincent, B.
\textit{Polymers at Interfaces}; Chapman \& Hall: London, 1993.

\bibitem{kawakatsu book}
Kawakatsu, T.
\textit{Statistical Physics of Polymers}; Splinger-Verlag: Berlin, 2004.

\bibitem{Laradji}
Laradji, M.; Shi, A.-C.; Noolandi, J.; Desai, C. R.
\textit{Macromolecules} \textbf{1997}, 30, 3242-3255.

\bibitem{MatsenComment}
Matsen, M.W.
\textit{Macromolecules} \textbf{1998}, 80, 201.

\bibitem{Matsen Cylinder}
Matsen, M. W.
\textit{J. Chem. Phys.} \textbf{2001}, 114, 8165-8173.
\bibitem{Matsen Gyroid}
Matsen, M. W.
\textit{Phys. Rev. Lett.} \textbf{1998}, 80, 4470-4473.

\bibitem{Fraaije}
Fraaije, J. G. E. M.
\textit{J. Chem. Phys.} \textbf{1993}, 99, 9202-9212.

\bibitem{Zvelindovsky}
Zvelindovsky, A. V.; Sevink, G. J. A.; van Vlimmeren, B. A. C.; Maurits, N. M.; Fraaije, J. G. E. M.
\textit{Phys. Rev. E} \textbf{1998}, 57, R4879-R4882.

\bibitem{Hasegawa Doi}
Hasegawa, R.; Doi, M.
\textit{Macromolecules} \textbf{1997}, 30, 5490-5493.

\bibitem{Hamley latest}
Hamley, I. W.; Castelletto, V.; Mykhaylyk, O. O.; Yang, Z.; May, R. P.; Lyakhova, K. S.; Sevink, G. J. A.; Zvelindovsky, A. V.
\textit{Langmuir} \textbf{2004}, 20, 10785-10790.

\bibitem{Morita Kawakatsu}
Morita, H.; Kawakatsu T.; Doi, M.
\textit{Macromolecules} \textbf{2001}, 34, 8777-8783.

\bibitem{Barrat}
Barrat, J. L.; Fredrickson, G. H.; Sides, S. W.
\textit{J. Phys. Chem.} \textbf{2005}, 109, 6694-6700.

\bibitem{SUSHI}
Honda, T.; et al. 
\textit{SUSHI Users Manual}, OCTA (http://octa.jp).

\bibitem{Andersen}
H. C. Andersen 
\textit{J. Chem. Phys.} \textbf{1980}, 72, 2384-2393.

\bibitem{Numerical Recipe in C}
Press, W. H.; Teukolsky, S. A.; Vetterling, W. T.; Flannery, B. P.
\textit{Numerical Recipes in C}; Cambridge University Press: Cambridge, 1992.

\bibitem{Computer Simulation of Liquids}
Allen, M. P.; Tildesley, D. J.
\textit{Computer simulation of liquids}; Oxfrd University Press: New York, 1987.


\end{thebibliography}
\end{document}